\documentclass[12pt,aps,showpacs,eqsecnum,nofootinbib,floatfix]{revtex4}
\usepackage{bbding}
\usepackage{pifont}
\usepackage{subfigure}
\usepackage{epsfig}
\usepackage{array}
\usepackage{amsmath}
\usepackage{amsfonts}
\usepackage{amssymb}
\usepackage{color}
\usepackage{graphicx}
\usepackage{mathrsfs}
\usepackage{multirow}

\def\be{\begin{equation}}
\def\ee{\end{equation}}
\def\bea{\begin{eqnarray}}
\def\eea{\end{eqnarray}}

\newcommand{\omits}[1]{}

\begin{document}

\title{Entropy of Higher Dimensional Charged Gauss-Bonnet Black hole in de Sitter Space}
\author{Xiong-Ying Guo$^{a,b}$, Huai-Fan Li$^{a,b}$\footnote{Email: huaifan.li@stu.xjtu.edu.cn; huaifan999@sxdtdx.edu.cn(H.-F. Li)} and Li-Chun Zhang$^{b}$}

\medskip

\affiliation{\footnotesize$^a$Department of Physics, Shanxi Datong
University,  Datong 037009, China\\
\footnotesize$^b$Institute of Theoretical Physics, Shanxi Datong
University, Datong 037009, China}

\begin{abstract}
The fundamental equation of the thermodynamic system gives the relation between
internal energy, entropy and volume of two adjacent equilibrium states. Taking higher
dimensional charged Gauss-Bonnet black hole in de Sitter
space as a thermodynamic system, the state parameters have to meet the fundamental equation of
thermodynamics. We introduce the effective thermodynamic quantities to describe the black hole in de Sitter space. Considering that in the lukewarm case the temperature of the black hole horizon is equal to that of the cosmological horizon, the effective temperature of spacetime is the same, we conjecture that the effective temperature has the same value. In this way, we can obtain the entropy formula of spacetime by solving the differential equation. We find that the total entropy contain an extra terms besides the sum of the entropies of the two horizons. The corrected terms of the entropy is a function of horizon radius ratio, and is independent of the charge of the spacetime.

\end{abstract}

 \maketitle

\section{Introduction}
There are black hole horizon and cosmological horizon for higher
dimensional charged Gauss-Bonnet(GB) black hole in de Sitter space. The thermodynamic quantities on the black hole horizon
and the cosmological horizon all satisfy the first law of thermodynamics, moreover the corresponding entropies both fulfill the
area formulae~\cite{Bekenstein72,Bekenstein74,Bekenstein73,Bardeen}.
In recent years, the investigation of black hole properties in de sitter space has
received a lot of attention~\cite{Dolan,Sekiwa,Kubiznak,Xu,Wang,Urano,Bhattacharya,McInerney,Guo,Guo16,
Zhao,Ma,Katsuragawa,Zhang,Li,Bhattacharya13,Kanti,Pappas,Pejhan,Cuyubamba,Konoplya,Cai02,Cai}.
However, in general the radiation temperatures corresponding to the two horizons
are not equal. For this reason, when taking the higher
dimensional charged GB black hole in de Sitter space as a thermodynamic system,
the system is usually unstable. The thermodynamic quantities on the black hole horizon and
the cosmological one in de Sitter space are functions of mass $M$, charge $Q$ and cosmological constant
$\Lambda$ respectively. These quantities, which correspond to the different horizons, are not
independent of each other. Considering the relation between the thermodynamic quantities on
the two horizons is very important for studying the thermodynamic properties of de Sitter space.

Considering the relation between the black hole horizon and the cosmological one and the higher dimensional charged GB black hole in de Sitter space as a thermodynamics system,
we obtain the effective thermodynamics quantifies of the higher dimensional charged GB black hole in de Sitter space.
In the lukewarm case, the temperature of the black hole horizon and that of the cosmological horizon are the same. we conjecture
that the effective temperature should also take the same value in the special case. In this way,
we provide the differential equation which the entropy
of both horizons satisfies. We assume that the total entropy includes the sum of both horizons entropy and the interaction terms.
For the entropy corresponding to the two horizons is a function of horizon radius and the
effective GB coefficient $\tilde {\alpha}$, the interaction terms of the corrected
entropy is a function of horizon radius and the effective GB coefficient
$\tilde {\alpha}$ too, and is independent of the charge of the spacetime. The result we obtained is self-consistent.
This work, which includes that constructing an self-consistent formula for the thermodynamics quantities of de Sitter
spacetime and studying the stability and the phase transition of de Sitter space, can provide more message
for the gravity theory of the de Sitter space. The issue can help that we have a more clear understanding for the
classics and quantities properties of de Sitter space. So, the issue which study the phase transition and critical
behavior of de Sitter space is worth thinking and studying.

This paper is organized as follows. In Sec.\ref{re} we simple review the thermodynamics
quantities of black hole horizon and cosmological horizon of the higher dimensional charged
GB black hole in de Sitter space, obtain the
condition that the effective temperature of the horizon approaches to zero.
In the base that the higher dimensional charged GB black hole in de Sitter space
 satisfies the first laws of thermodynamics, the
entropy and the effect temperature of higher dimensional charged GB black hole in de Sitter space is obtained, we study the
condition that higher dimensional charged GB black hole in de Sitter space satisfies the stable equilibrium of the thermodynamics
in Sec.\ref{eff}. Sec.\ref{con} is devoted to conclusions.(we
use the units $G_{n + 1} = \hbar = k_B = c = 1)$

\section{Charged GB Black Hole in de Sitter space}
\label{re}

Higher derivative curvature terms occur in many occasions, such as in the
semiclassically quantum gravity and in the effective low-energy action of
superstring theories. Among the many theories of gravity with higher
derivative curvature terms, due to the special features the GB
gravity has attract much interest. The thermodynamic properties and
phase structures of GB-AdS black hole have been briefly discussed in~\citep{Cai,nojiri,nojiri2,nojiri3}.
In Refs.~\cite{Hendi,Hendi2,Hendi3,Hendi4,Wei,Cai13,Xu14,Wen}, the critical phenomena and phase transition of the charged
GB-AdS black hole have been studied extensively. In this paper, we study the
thermal properties of charged GB-dS black hole after considering the
connections between the black hole horizon and the cosmological horizon~\cite{Ma14}.

The action of $d$-dimensional Einstein-GB-Maxwell theory with
a bare cosmological constant $\Lambda $ reads~\cite{Cai13}
\begin{equation}
\label{eq1}
I = \frac{1}{16\pi }\int {d^dx\sqrt { - g} } \left[ {R - 2\Lambda + \alpha
(R_{\mu \nu \gamma \delta } R^{\mu \nu \gamma \delta } - 4R_{\mu \nu }
R^{\mu \nu } + R^2 - 4\pi F_{\mu \nu } F^{\mu \nu }} \right],
\end{equation}
where the GB coupling $\alpha $ has dimension [length]$^{2}$ and can be
identified with the inverse string tension with positive value if the
theory is incorporated in string theory~\cite{Kubiznak12}, thus we shall consider only the
case $\alpha > 0$. $F_{\mu \nu } $ is the Maxwell field strength defined as
$F_{\mu \nu } = \partial _\mu A_\mu - \partial _\nu A_\nu $ with vector
potential $A_\mu $. In addition, let us mention here that the GB term is a
topological term in $d = 4$ dimensions and has no dynamics in this case.
Therefore we will consider $d \ge 5$ in what follows.
\begin{equation}
\label{eq2}
ds^2 = - f(r)dt^2 + f^{ - 1}(r)dr^2 + r^2h_{ij} dx^idx^j,
\end{equation}
where $h_{ij} dx^idx^j$ represent the line element of a ($d-2$)-dimensional maximal
symmetric Einstein space with constant curvature $(d - 2)(d - 3)k$ and
volume $\Sigma _k $. Without loss of the generality, one may take $k = 1,0$
and $ - 1$, corresponding to the spherical, Ricci fiat and hyperbolic
topology of the black hole horizon, respectively. The metric function $f(r)$
is given by~\cite{Hendi,Wei,Cai13,Xu14,Wen}
\begin{equation}
\label{eq3}
f(r) = k + \frac{r^2}{2\tilde {\alpha }}\left[ {1 - \sqrt {1 + \frac{64\pi
\tilde {\alpha }M}{(d - 2)\Sigma _k r^{d - 1}} - \frac{2\tilde {\alpha
}Q^2}{(d - 2)(d - 3)r^{2d - 4}} + \frac{8\tilde {\alpha }\Lambda }{(d - 1)(d
- 2)}} } \right],
\end{equation}
where $\tilde {\alpha } = (d - 3)(d - 4)\alpha $, $M$ and $Q$ are the mass
and charge of black hole respectively, and pressure $P$
\begin{equation}
\label{eq4}
P = - \frac{\Lambda }{8\pi } = \frac{(d - 1)(d - 2)}{16\pi l^2}.
\end{equation}
In order to have a well-defined vacuum solution with $M = Q = 0$,
the effective GB coefficient $\tilde {\alpha }$ and pressure $P$
have to satisfy the following constraint
\begin{equation}
\label{eq5}
0 < \frac{64\pi \tilde {\alpha }P}{(d - 1)(d - 2)} \le 1.
\end{equation}
The black hole horizon $r_+$ and cosmological horizon $r_c$ satisfies the equation $f(r_{ + ,c} ) = 0$. The
equations $f(r_ + ) = 0$ and $f(r_c ) = 0$ are rearranged to
\begin{equation}
\label{eq6}
M = \frac{(d - 2)\Sigma _k r_ + ^{d - 3} }{16\pi }\left( {k +
\frac{k^2\tilde {\alpha }}{r_ + ^2 }} \right) - \frac{\Sigma _k r_ + ^{d -
1} \Lambda }{8\pi (d - 1)} + \frac{\Sigma _k Q^2}{32\pi (d - 3)r_ + ^{d - 3}
}
\end{equation}
\begin{equation}
\label{eq7}
M = \frac{(d - 2)\Sigma _k r_c^{d - 3} }{16\pi }\left( {k + \frac{k^2\tilde
{\alpha }}{r_c^2 }} \right) - \frac{\Sigma _k r_c^{d - 1} \Lambda }{8\pi (d
- 1)} + \frac{\Sigma _k Q^2}{32\pi (d - 3)r_c^{d - 3} }
\end{equation}
From Eqs.(\ref{eq6}) and (\ref{eq7}), we can obtain
\[
M = \frac{(d - 2)\Sigma _k r_c^{d - 3} x^{d - 3}}{16\pi }\frac{(1 - x^2)}{(1
- x^{d - 1})}\left( {k + \frac{k^2\tilde {\alpha }_c }{x^2}(1 + x^2)}
\right)
\]
\[
 + \frac{\Sigma _k Q^2}{32\pi (d - 3)r_c^{d - 3} (1 - x^{d - 1})x^{d -
3}}\left( {1 - x^{2d - 4}} \right)
\]
\begin{equation}
\label{eq8}
 = \frac{r_c^2 \Sigma _k }{16\pi }f_{H1} (x,\tilde {\alpha }_c ) +
\frac{Q^2}{r_c^2 }\frac{\Sigma _k }{16\pi }f_{H2} (x),
\end{equation}
here
\[
f_{H1} = (d - 2)x^{d - 3}\frac{(1 - x^2)}{(1 - x^{d - 1})}\left( {k +
\frac{k^2\tilde {\alpha }_c }{x^2}(1 + x^2)} \right),
\]
\begin{equation}
\label{eq9}
f_{H2} = \frac{1}{2(d - 3)(1 - x^{d - 1})x^{d - 3}}\left( {1 - x^{2d - 4}}
\right).
\end{equation}
\begin{equation}
\label{eq10}
\Lambda = \frac{(d - 1)(d - 2)}{2r_c^2 (1 - x^{d - 1})}\left( {k(1 - x^{d -
3}) + \frac{\tilde {\alpha }k^2}{r_c^2 }(1 - x^{d - 5})} \right) - \frac{(d
- 1)Q^2(1 - x^{d - 3})}{4(d - 3)r_c^{2d - 4} x^{d - 3}(1 - x^{d - 1})}.
\end{equation}
here $x = r_ + / r_c $, $\tilde {\alpha }_c = \tilde {\alpha } / r_c^2 $.

From Eqs.(\ref{eq3}), (\ref{eq8}) and (\ref{eq10}), we can obtain
\[
f'(r_ + ) = - \frac{2k}{r_ + }
 + \frac{(d - 1)kr_ + r_c^{d - 3} (r_c^2 - r_ + ^2 )}{(r_ + ^2 + 2\alpha
k)(r_c^{d - 1} - r_ + ^{d - 1} )}\left( {1 + \frac{\alpha k(r_c^2 + r_ + ^2
)}{r_ + ^2 r_c^2 }} \right)
\]

\begin{equation}
\label{eq11}
 - \frac{Q^2}{2(r_ + ^2 + 2\alpha k)(d - 2)r_ + ^{2d - 7} }\left[ {1 -
\frac{(d - 1)r_ + ^{d - 1} (r_c^{d - 3} - r_ + ^{d - 3} )}{(d - 3)r_c^{d -
3} (r_c^{d - 1} - r_ + ^{d - 1} )}} \right]
 = A - Q^2B
\end{equation}
\[
f'(r_c ) = - \frac{2k}{r_c }
 + \frac{(d - 1)kr_c r_ + ^{d - 3} (r_c^2 - r_ + ^2 )}{(r_c^2 + 2\alpha
k)(r_c^{d - 1} - r_ + ^{d - 1} )}\left( {1 + \frac{\alpha k(r_c^2 + r_ + ^2
)}{r_ + ^2 r_c^2 }} \right)
\]
\begin{equation}
\label{eq12}
 - \frac{Q^2}{2(r_c^2 + 2\alpha k)(d - 2)r_c^{2d - 7} }\left[ {1 - \frac{(d
- 1)r_c^{d - 1} (r_c^{d - 3} - r_ + ^{d - 3} )}{(d - 3)r_ + ^{d - 3} (r_c^{d
- 1} - r_ + ^{d - 1} )}} \right]
 = C - Q^2D
\end{equation}
here
\[
A = \frac{1}{r_c }f_A (x,\tilde {\alpha }_c ) = - \frac{2k}{r_c x} +
\frac{(d - 1)kx(1 - x^2)}{r_c (x^2 + 2\tilde {\alpha }_c k)(1 - x^{d -
1})}\left( {1 + \frac{\tilde {\alpha }_c k(1 + x^2)}{x^2}} \right),
\]

\[
B = \frac{1}{r_c^{2d - 5} }f_B (x,\tilde {\alpha }_c ) = \frac{1}{2(d -
2)(x^2 + 2\tilde {\alpha }_c k)r_c^{2d - 5} x^{2d - 7}}\left( {1 - \frac{(d
- 1)x^{d - 1}(1 - x^{d - 3})}{(d - 3)(1 - x^{d - 1})}} \right),
\]

\[
C = \frac{1}{r_c }f_C (x,\tilde {\alpha }_c ) = - \frac{2k}{r_c } + \frac{(d
- 1)kx^{d - 3}(1 - x^2)}{r_c (1 + 2\tilde {\alpha }_c k)(1 - x^{d -
1})}\left( {1 + \frac{\tilde {\alpha }_c k(1 + x^2)}{x^2}} \right)
\]
\begin{equation}
\label{eq13}
D = \frac{1}{r_c^{2d - 5} }f_D (x,\tilde {\alpha }_c ) = \frac{1}{2(d - 2)(1
+ 2\tilde {\alpha }_c k)r_c^{2d - 5} }\left( {1 - \frac{(d - 1)(1 - x^{d -
3})}{(d - 3)x^{d - 3}(1 - x^{d - 1})}} \right).
\end{equation}
Some thermodynamic quantities associated with the cosmological horizon are
\begin{equation}
\label{eq14}
T_c = - \frac{f'(r_c )}{4\pi } = - \frac{C - Q^2D}{4\pi },
\quad
S_c = \frac{\Sigma _k r_c^{d - 2} }{4}\left( {1 + \frac{2\tilde {\alpha }_c
k(d - 2)}{(d - 4)}} \right),
V_c = \frac{\Sigma _k r_c^{d - 1} }{d - 1}.
\end{equation}
$T_c$, $S_c $ and $V_c $ denote the Hawking temperature, the entropy and
the volumes.

\begin{figure}[!htbp]
\center{
\includegraphics[width=7cm,keepaspectratio]{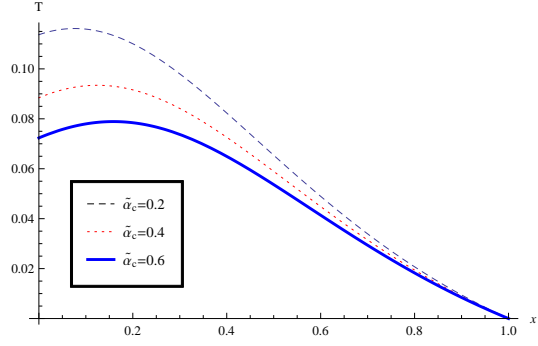}\hspace{0.5cm}}\\
\caption{$T$ with respect to $x$. we set $r_c=1$.\label{Tx}}
\end{figure}

For the black hole horizon, associated thermodynamic quantities are
\begin{equation}
\label{eq15}
T_ + = \frac{f'(r_ + )}{4\pi } = \frac{A - Q^2B}{4\pi r},
\quad
S_ + = \frac{\Sigma _k r_c^{d - 2} x^{d - 2}}{4}\left( {1 + \frac{2\tilde
{\alpha }_c k(d - 2)}{(d - 4)x^2}} \right),
V_ + = \frac{\Sigma _k r_c^{d - 1} x^{d - 1}}{d - 1}.
\end{equation}
From Eqs.(\ref{eq10}) and (\ref{eq11}), when the charge $Q$ of the spacetime satisfies
\begin{equation}
\label{eq16}
Q^2 = \frac{A + C}{B + D} = r_c^{2d - 6} \frac{f_A (x,\tilde {\alpha }_c ) +
f_C (x,\tilde {\alpha }_c )}{f_B (x,\tilde {\alpha }_c ) + f_D (x,\tilde
{\alpha }_c )},
\end{equation}
the temperature of the black hole horizon and the ones of the cosmological horizon
is equal,
\begin{equation}
\label{eq17}
T = T_ + = T_c = \frac{1}{4\pi }\frac{f_A (x,\tilde {\alpha }_c )f_D
(x,\tilde {\alpha }_c ) - f_C (x,\tilde {\alpha }_c )f_B (x,\tilde {\alpha
}_c )}{r_c [f_B (x,\tilde {\alpha }_c ) + f_D (x,\tilde {\alpha }_c )]}.
\end{equation}
From Eq.(\ref{eq17}), we plot the $T-x$ plane with different $\tilde {\alpha }_c $, $d=5$ and $k=1$ in Fig.\ref{Tx}.

\section{THE EFFECTIVE THERMODYNAMICS of HGBDS}
\label{eff}

\begin{figure}[!htbp]
\center{ \subfigure[]{ \label{1-a}
\includegraphics[width=7cm,keepaspectratio]{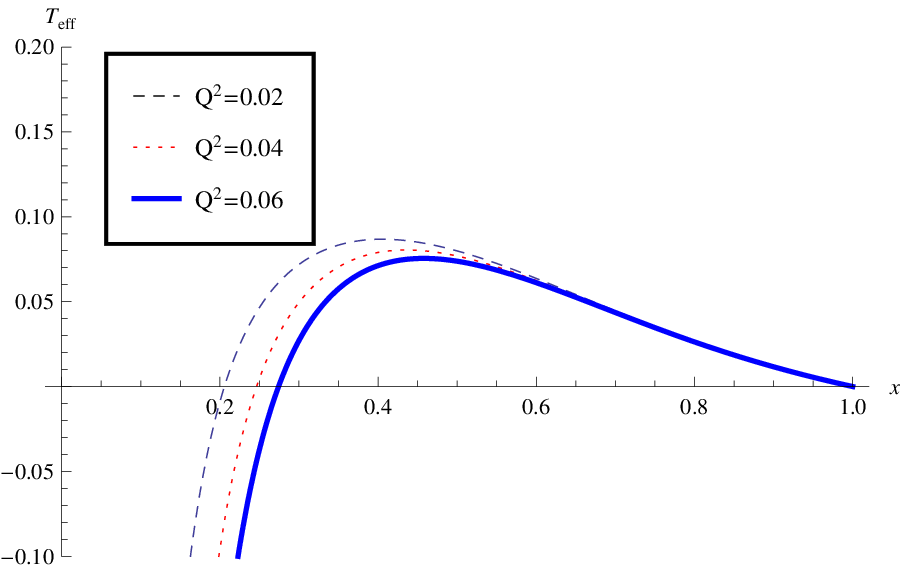}\hspace{0.5cm}}
\subfigure[]{ \label{1-b}
\includegraphics[width=7cm,keepaspectratio]{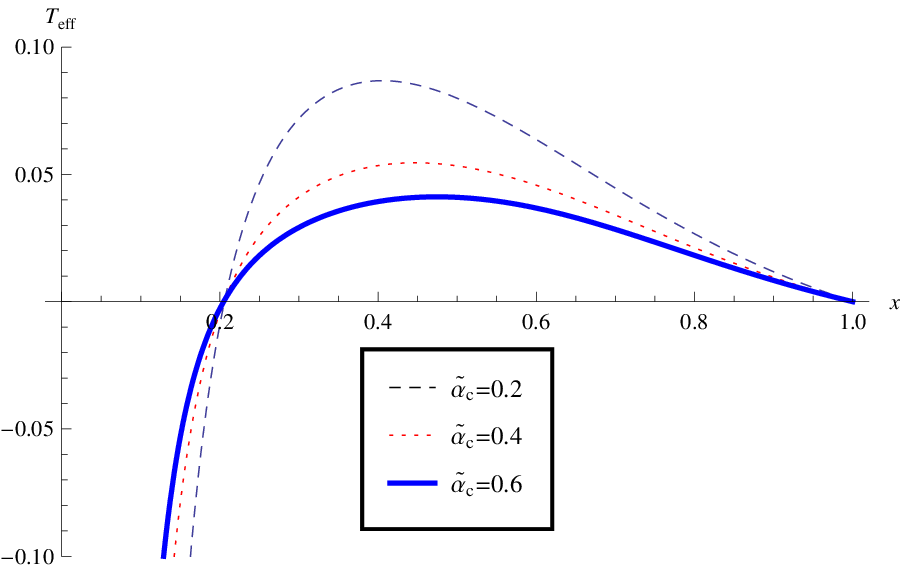}}
\caption {the effective temperature $T_{eff}$ versus $x$. In (a), $T_{eff}-x$ with different $Q$ and fixed $\tilde\alpha_c=0.2$. In (b), $T_{eff}-x$ with different $\tilde\alpha_c$ and fixed $Q^2=0.02$.  we set $r_c=1$.\label{fx}}
}
\end{figure}

\begin{figure}[!htbp]
\center{
\includegraphics[width=7cm,keepaspectratio]{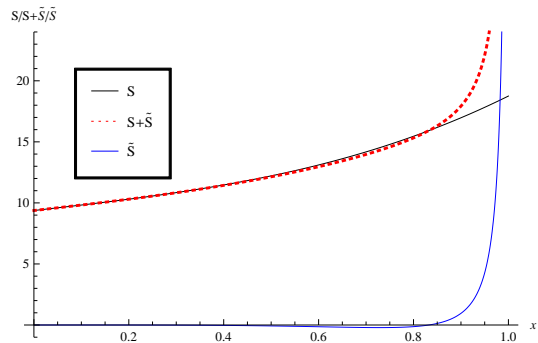}\hspace{0.5cm}}\\
\caption {The uncorrected entropy $S$, the corrected terms $\tilde{S}$ for entropy and  the total entropy $S+\tilde{S}$ with respect to $x$. we set $r_c=1$.\label{Sx}}
\end{figure}

Considering the connection between the black hole horizon and the
cosmological horizon, we can derive the effective thermodynamic quantities
and corresponding the first law of black hole thermodynamics~\cite{Dolan,Zhang,Li}
\begin{equation}
\label{eq18}
dM = T_{eff} dS - P_{eff} dV + \varphi _{eff} dQ,
\end{equation}
here the thermodynamic volume is that between the black hole horizon and the
cosmological horizon, namely~\cite{Dolan,Ma14}
\begin{equation}
\label{eq19}
V = V_c - V_ + = \frac{\Sigma _k }{d - 1}\left( {r_c^{d - 1} - r_ + ^{d - 1}
} \right) = \frac{\Sigma _k }{d - 1}r_c^{d - 1} \left( {1 - x^{d - 1}}
\right),
\end{equation}
Considering the formula of the entropy for both horizons, dimension and the terms $\tilde {\alpha }_c
,k$, we take the total entropy as
\begin{equation}
\label{eq20}
S = \frac{\Sigma _k r_c^{d - 2} }{4}\left[ {f(x) + \tilde {\alpha }_c kf_1
(x) + \tilde {\alpha }_c^2 k^2f_2 (x)} \right],
\end{equation}
\[
S = S_ + + S_c + \tilde {S} = \frac{\Sigma _k r_ + ^{d - 2} }{4}\left( {1 +
\frac{2(d - 2)\tilde {\alpha }k}{(d - 4)r_ + ^2 }} \right) + \frac{\Sigma _k
r_c^{d - 2} }{4}\left( {1 + \frac{2(d - 2)\tilde {\alpha }k}{(d - 4)r_c^2 }}
\right) + \tilde {S}
\]
\[
 = \frac{\Sigma _k }{4}\left( {r_c^{d - 2} + r_ + ^{d - 2} } \right) +
\frac{\Sigma _k }{2}\frac{(d - 2)\tilde {\alpha }k}{(d - 4)}\left( {r_c^{d -
4} + r_ + ^{d - 4} } \right) + \tilde {S}
\]
\begin{equation}
\label{eq21}
 = \frac{\Sigma _k }{4}r_c^{d - 2} \left( {1 + x^{d - 2}} \right) +
\frac{\Sigma _k }{2}\frac{(d - 2)\tilde {\alpha }_c k}{(d - 4)}r_c^{d - 4}
\left( {1 + x^{d - 4}} \right) + \tilde {S}.
\end{equation}
Taking
\[
\tilde {S} = \frac{\Sigma _k r_c^{d - 2} }{4}f(x,\tilde {\alpha }_c )
\]
\begin{equation}
\label{eq22}
S = \frac{\Sigma _k }{4}r_c^{d - 2} \left[ {\left( {1 + x^{d - 2} +
f(x,\tilde {\alpha }_c )} \right) +\frac{2(d - 2)\tilde {\alpha }_c k}{(d -
4)r_c^2}\left( {1 + x^{d - 4}} \right)} \right].
\end{equation}
Here the undefined function $f(x)$, $f_1 (x)$ and $f_2 (x)$ represents the extra
contribution from the correlations of the two horizons. When taking $Q$, $\tilde
{\alpha }_c$ as constant, substituting Eqs.(\ref{eq8}),(\ref{eq19}) and (\ref{eq20})
into Eq. (\ref{eq18}), one obtain the effective temperature $T_{eff} $
\begin{equation}
\label{eq23}
T_{eff} = \frac{T_1 (x)}{4\pi r_c T_2 (x)},
\end{equation}
here
\[
T_1 (x) = \left( {1 - x^{d - 1}} \right)\left[ {f_{H1} '(x,\tilde {\alpha
}_c ) + \frac{Q^2}{r_c^{2d - 6} }f_{H2} '(x)} \right] + x^{d - 2}(d -
3)\left[ {f_{H1} (x,\tilde {\alpha }_c ) - \frac{Q^2}{r_c^{2d - 6} }f_{H2}
(x)} \right],
\]
\[
T_2 (x) = (1 - x^4)\left[ {f'(x) + \tilde {\alpha }_c kf_1 '(x) + \tilde
{\alpha }_c^2 k^2f_2 '(x)} \right]
\]
\begin{equation}
\label{eq24}
 + 3x^3\left[ {f(x) + \tilde {\alpha }_c kf_1 (x) + \tilde {\alpha }_c^2
k^2f_2 (x)} \right].
\end{equation}
When $Q$ satisfies Eq.(\ref{eq16}),
the temperature of both horizons is equal. In this case we think
that the effective temperature of spacetime is the radiation temperature
\begin{equation}
\label{eq25}
T_{eff} = \frac{\tilde {T}_1 (x)}{4\pi r_c T_2 (x)} = \frac{1}{4\pi
}\frac{f_A (x,\tilde {\alpha }_c )f_D (x,\tilde {\alpha }_c ) - f_C
(x,\tilde {\alpha }_c )f_B (x,\tilde {\alpha }_c )}{r_c [f_B (x,\tilde
{\alpha }_c ) + f_D (x,\tilde {\alpha }_c )]},
\end{equation}
here
\[
\tilde {T}_1 (x) = \left( {1 - x^{d - 1}} \right)\left[ {f_{H1} '(x,\tilde
{\alpha }_c ) + f_{H2} '(x)\frac{f_A (x,\tilde {\alpha }_c ) + f_C (x,\tilde
{\alpha }_c )}{f_B (x,\tilde {\alpha }_c ) + f_D (x,\tilde {\alpha }_c )}}
\right]
\]
\begin{equation}
\label{eq26}
 + (d - 3)x^{d - 2}\left[ {f_{H1} (x,\tilde {\alpha }_c ) - f_{H2}
(x)\frac{f_A (x,\tilde {\alpha }_c ) + f_C (x,\tilde {\alpha }_c )}{f_B
(x,\tilde {\alpha }_c ) + f_D (x,\tilde {\alpha }_c )}} \right].
\end{equation}
From(\ref{eq25}), one get
\begin{equation}
\label{eq27}
T_2 (x) = \frac{\tilde {T}_1 (x)[f_B (x,\tilde {\alpha }_c ) + f_D (x,\tilde
{\alpha }_c )]}{f_A (x,\tilde {\alpha }_c )f_D (x,\tilde {\alpha }_c ) - f_C
(x,\tilde {\alpha }_c )f_B (x,\tilde {\alpha }_c )}.
\end{equation}
When $d = 5$, from Eq.(\ref{eq27}) one obtain
\[
T_2 (x) = 3\frac{x^2(1 + x^5) + 2\tilde {\alpha }_c k(1 + x^7)}{(1 - x^4)}
\]
\begin{equation}
\label{eq28}
 + 3kx^3\tilde {\alpha }_c \frac{[1 + x + 3x^2 + x^3 + x^4 + 2\tilde {\alpha
}_c k(1 - x + x^2)]}{(1 + x)(1 - x^2)}.
\end{equation}
Substituting Eq.(\ref{eq24}) into Eq.(\ref{eq28}), comparisons with the power of $\tilde {\alpha }_c$
the two sides of this equation, one can obtain $f(x)$, $f_1(x)$ and $f_2(x)$ satisfies
\[
(1 - x^4)f'(x) + 3x^3f(x) = A_0 ,
\]
\[
(1 - x^4)f_1 '(x) + 3x^3f_1 (x) = A_1 ,
\]
\begin{equation}
\label{eq29}
(1 - x^4)f_2 '(x) + 3x^3f_2 (x) = A_2 ,
\end{equation}
respectively, here
\[
A_0 = \frac{3 x^2(1 + x^5)}{(1 - x^4)},
\]
\[
A_1 = \frac{3(2 + 2x + x^3 + x^4 + 4x^5 + 2x^6 + 6x^7 + 3x^8 + x^9)}{(1 + x)(1
- x^4)},
\]
\begin{equation}
\label{eq30}
A_2 = \frac{6x^3(1 - x + x^2)}{(1 + x)(1 - x^2)} = \frac{6x^3(1 + x^3)}{(1 +
x)^2(1 - x^2)}.
\end{equation}
When $x \to 0$, there are only cosmological horizon in de Sitter space. From Eq.(\ref{eq14}), we take the initial value $f(0) = 1,f_1 (0) = 6, f_2 (0) = 0$,
\[
f(x) = \frac{11}{7}(1 - x^4)^{3 / 4} - \frac{4(1 + x^4)(1 - x + x^2) -
11x^3}{7(1 - x)(1 + x^2)}
\]
\begin{equation}
\label{eq31}
 = \frac{11}{7}(1 - x^4)^{3 / 4} - \frac{11 - x(11 + 3x^3)(1 - x + x^2)}{7(1
- x)(1 + x^2)} + 1 + x^3
 = \tilde {f}(x) + 1 + x^3.
\end{equation}
\[
f_1 (x) = \frac{1975}{308}(1 - x^4)^{3 / 4} + \frac{1}{77(1 + x)^2(1 - x)(1
+ x^2)}\left( { - 127 + 76} \right.x + 211x^2 + 8x^3
\]
\[
 + 280x^4 + 140x^5 + 2x^6 + 76x^7 + 60x^8
\]
\[
 + 259x(1 - x^4)^{3 / 4}(1 + x)(1 - x^4) _2F_1\left[
{\textstyle{1 \over 4},\textstyle{3 \over 4},\textstyle{5 \over 4},x^4}
\right]
\]
\[
\left. { - 8x^2(1 - x^4)^{3 / 4}(1 + x)(1 - x^4) _2F_1\left[
{\textstyle{1 \over 2},\textstyle{3 \over 4},\textstyle{3 \over 2},x^4}
\right]} \right)
\]
\[
 = \frac{1975}{308}(1 - x^4)^{3 / 4} + \frac{1}{77(1 + x)^2(1 - x)(1 +
x^2)}\left( { - 204 - 68} \right.x + 134x^2 + 8x^3
\]
\[
 + 357x^4 + 284x^5 + 79x^6 + 76x^7 + 60x^8 + 1 + x
\]
\[
 + 259x(1 - x^4)^{3 / 4}(1 + x)(1 - x^4) _2F_1\left[
{\textstyle{1 \over 4},\textstyle{3 \over 4},\textstyle{5 \over 4},x^4}
\right]
\]
\[
\left. { - 8x^2(1 - x^4)^{3 / 4}(1 + x)(1 - x^4) _2F_1\left[
{\textstyle{1 \over 2},\textstyle{3 \over 4},\textstyle{3 \over 2},x^4}
\right]} \right)
\]
\begin{equation}
\label{eq32}
 = \tilde {f}_1 (x) + 6(1 + x).
\end{equation}
we can obtain $f_2(x)$ with the  numerical calculation. From Eqs.(\ref{eq14}) and (\ref{eq15}), we obtain the entropy of black hole
horizon and cosmological horizon as
\begin{equation}
\label{eq33}
S = \frac{\Sigma _k }{4}r_c^{d - 2} \left[ {1 + x^{d - 2} + 2\frac{(d -
2)\tilde {\alpha }_c k}{(d - 4)}\left( {1 + x^{d - 4}} \right)} \right].
\end{equation}
Comparing Eqs.(\ref{eq31}), (\ref{eq32}) and (\ref{eq33}), we can obtain the corrected terms of the
system entropy form the interaction of both horizons.
\begin{equation}
\label{eq34}
\tilde {S} = \tilde {f}(x) + \tilde {\alpha }_c k \tilde f_1
(x) + \tilde {\alpha }_c^2 k^2 f_2 (x).
\end{equation}
Substituting Eq.(\ref{eq28}) into Eq.(\ref{eq23}), we can plot the $T_{eff} (x) - x$ plane with
different $Q,\tilde {\alpha }_c$ and $r_c = 1$, $k=1$ in Fig.\ref{fx}.
From Eq.(\ref{eq20}), we can plot the $S-x$, $\tilde{S}-x$ and $(\tilde{S}+S)-x$ with $\tilde {\alpha }_c=0.15$ and $k=1$ in Fig.\ref{Sx}.
The specific heat of higher dimensional charged GB black hole in de Sitter space can be defined as
\begin{equation}
\label{eq36}
C_{Q,\tilde{\alpha}_c} = T_{eff} \left( {\frac{\partial S}{\partial T_{eff} }}
\right)_{Q,\tilde{\alpha}_c} .
\end{equation}

\begin{figure}[!htbp]
\center{ \subfigure[]{ \label{1-a}
\includegraphics[width=7cm,keepaspectratio]{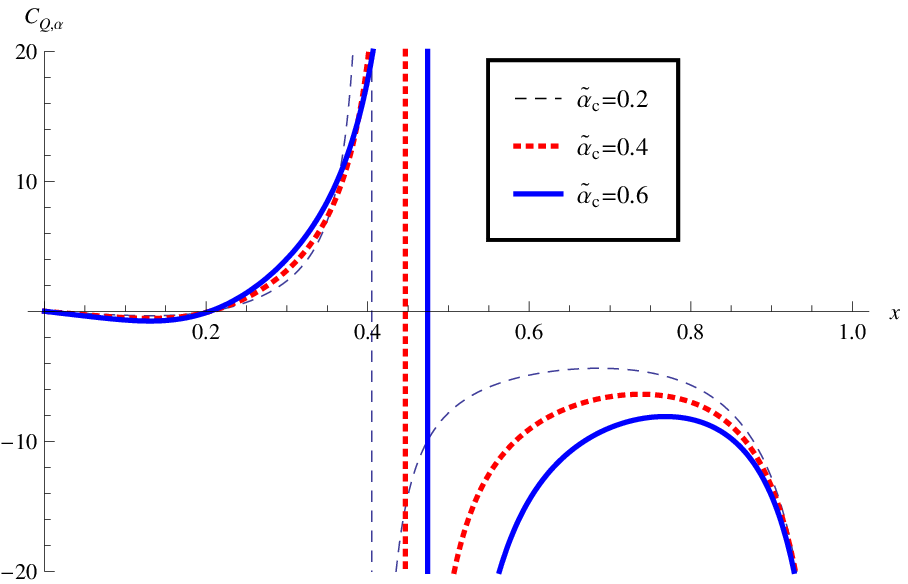}\hspace{0.5cm}}
\subfigure[]{ \label{1-b}
\includegraphics[width=7cm,keepaspectratio]{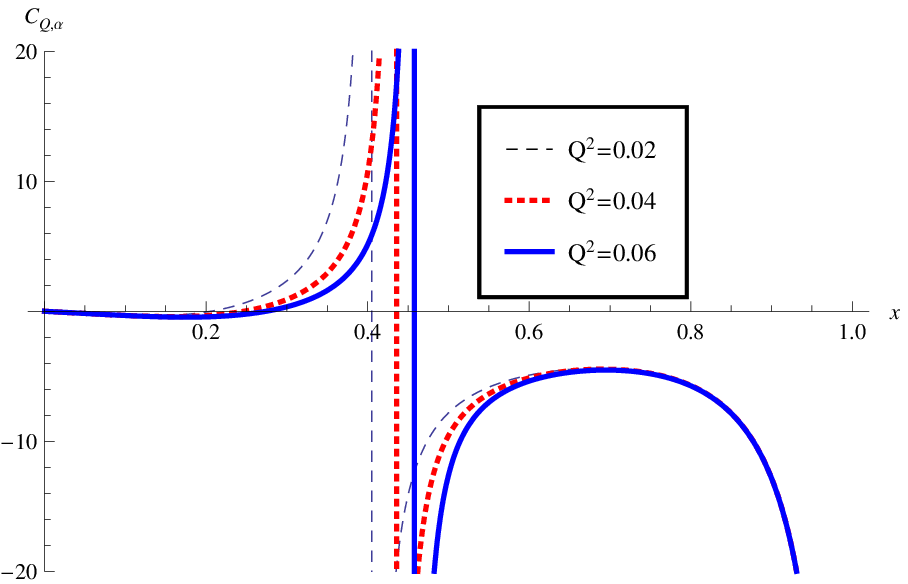}}
\caption {$C_{Q,\tilde{\alpha}_c}$ with respect to $x$. In (a), $C_{Q,\tilde{\alpha}_c}-x$ with different $Q^2$ and fixed $\tilde\alpha=0.2$. In (b), $C_{Q,\tilde{\alpha}_c}-x$ with different $\alpha$ and fixed $Q^2=0.02$.  we set $r_c=1$.\label{cqx}}
}
\end{figure}

From Fig.\ref{cqx}, when $x_0 <x<x_c$, the specific heat of system is positive, while $x_c<x<1$ or $0<x<x_0$, it is negative. This
means that the higher dimensional charged GB black hole in de Sitter space with $x_0 <x<x_c$ is thermodynamically stable. From the Fig.\ref{cqx},
we can find that the region of stable state of higher dimensional charged GB black hole in de Sitter space
is related to the charge $Q$ and the effective GB coefficient $\tilde\alpha$, i.e., the region of stable state becomes bigger as the effective GB coefficient $\tilde\alpha$ or the charge $Q$ is increases.

In this letter, we have presented the entropy of higher dimensional charged GB black hole in de Sitter space. It is not only the sum of the entropies of black hole horizon and the cosmological horizon, but also with an extra term from the correlation between the two horizons. This idea has twofold advantages. First, if without the extra term in the total entropy, the effective temperature is not the same as that of the black hole horizon and the cosmological horizon in the lukewarm case. This is not satisfactory. Second, the method of effective first law of thermodynamics lacks physical explanation or motivation. While taking advantage of the method, we obtain the corrected entropy of higher dimensional charged GB black hole in de Sitter space, which may make the method more acceptable.

\section{discussions and conclusions}
\label{con}
The thermodynamic quantities on the black hole horizon and the cosmological one in de Sitter space are functions of mass $M$, charge $Q$ and cosmological constant $\Lambda$ respectively. These quantities, which correspond to the different horizons, are not independent of each other. It is not possible to fully realize the thermodynamic properties of de Sitter space time by studying the thermodynamics system corresponding the two horizons in de Sitter space separately. So, we can take the states parameters in de Sitter space as the states parameters in the whole system. We know that the states parameters in the whole system satisfies the first thermodynamics laws. So, we can obtain the effective temperature $T_{eff}$ and the total entropy $S+\tilde S$ by the above discussion and calculation.

Without considering the correlation between the black hole horizon and cosmological horizon, we take the two horizons as the independent thermodynamics system respectively. Due the radiation temperature of the two horizon is different, the spacetime did not meet the requirements of the stability of the thermodynamic equilibrium, so the system is unstable. Considering the correlation between the black hole horizon and the cosmological horizon, we can obtain the only effective temperature $T_{eff}$ of the higher dimensional charged GB black hole in de Sitter space from Eq.(\ref{eq23}). From $C_{Q,\tilde{\alpha}_c} - x$ (Fig.\ref{cqx}), when $x > x_c $ or $x < x_0
$, we know that higher dimensional charge GB black hole in de Sitter space is unstable. The system cannot stable for the thermodynamic system that does not meet the thermodynamic equilibrium conditions. In the universe, there are only possible higher dimensional charged GB black hole in de Sitter space that meet the conditions of $x_0 < x < x_c$. Because the cosmological constant is associated with a vacuum that describes the most fundamental theories of nature, such as quantum gravity, and de Sitter space is closely connected with the evolution of our universe. A deeper understanding of the quantum nature of de Sitter is undoubtedly helpful to establish self-consistent quantum gravity theory.

\section*{Acknowledgements}
We would like to thank Prof. Zong-Hong Zhu and Meng-Sen Ma for their indispensable discussions and comments. This work was supported by the Young Scientists Fund of the National Natural Science Foundation of China (Grant No.11205097), in part by the National Natural Science Foundation of China (Grant No.11475108), Supported by Program for the Innovative Talents of Higher Learning Institutions of Shanxi, the Natural Science Foundation of Shanxi Province,China(Grant No.201601D102004) and the Natural Science Foundation for Young Scientists of Shanxi Province,China (Grant No.2012021003-4).

\end{document}